\documentclass{article}

\newcommand{\refeq}[1]{(\ref{#1})}

\newcommand{\refsec}[1]{Sect.~\ref{#1}}
\newcommand{\reffig}[1]{Fig.~\ref{#1}}

\newcommand{\opPi}{\mathbf{\Pi}}
\newcommand{\opE}{\mathbf{E}}
\newcommand{\opH}{\mathbf{H}}
\newcommand{\opD}{\mathbf{D}}
\newcommand{\opkappa}{\boldsymbol{\kappa}}
\newcommand{\vr}{\overline{V}}

\newcommand{\ud}{\mathrm{d}}

\usepackage{graphicx}
\usepackage{bm}
\usepackage{amssymb,amsmath,amsfonts}
\usepackage{epsfig}
\usepackage{enumerate}
\usepackage{epstopdf}
\usepackage[T1]{fontenc}
\usepackage{lmodern}
\usepackage{times}
\usepackage{psfrag}

\title{Ray and wave scattering in smoothly curved thin shell cylindrical ridges}

\author{Niels S\o ndergaard$^\mathrm{a}$ and David J. Chappell$^\mathrm{b}$\vspace{2mm}\\
$^\mathrm{a}$ inuTech GmbH, F\"{u}rther Stra{\ss}e 212, 90429 Nuremberg, Germany.\vspace{1mm}\\
$^\mathrm{b}$ School of Science and Technology, Nottingham Trent
University,\\ Nottingham NG11 8NS, United Kingdom.}
\date{}
\begin{document}
\maketitle
\begin{abstract}
We propose wave and ray approaches for modelling mid- and high-
frequency structural vibrations through smoothed joints on thin
shell cylindrical ridges. The models both emerge from a simplified
classical shell theory setting. The ray model is analysed via an
appropriate phase-plane analysis, from which the fixed points can be
interpreted in terms of the reflection and transmission properties.
The corresponding full wave scattering model is studied using the
finite difference method to investigate the scattering properties of
an incident plane wave. Through both models we uncover the
scattering properties of smoothed joints in the interesting
mid-frequency region close to the ring frequency, where there is a
qualitative change in the dynamics from anisotropic to simple
geodesic propagation.
\end{abstract}

\section{Introduction}

Thin shell components can be found in many large built-up mechanical
structures such as cars, ships, and aeroplanes. The prediction of
the mid- and high- frequency vibrational properties of these
structures becomes computationally prohibitive for standard
element-based methods, such as the finite element method
\cite{OZ03}. The main reasons for this limitation are: firstly, very
fine meshes are required for an adequate representation of the
highly oscillatory wave solutions and the computational complexity
grows with frequency raised to the power of the dimension of the
space being modelled. Secondly, small uncertainties originating from
the manufacturing process lead to a much larger variability in
vibro-acoustic responses in the high frequency range \cite{LD95},
meaning that the response of any individual manufactured structure
is of less interest to computer aided engineering practitioners than
the average responses.

Methods such as statistical energy analysis (SEA) \cite{LD95} and
ray tracing \cite{AG89} are more commonly applied for modelling wave
problems at high frequencies. SEA has traditionally proved more
popular for structural vibration problems with low damping
\cite{AL15}, whereas ray tracing has found its niche in applications
where relatively few reflections need to be tracked in computer
graphics \cite{AG89}, room acoustics \cite{HK00} and seismology
\cite{VC01}. Ray tracing has also been applied to elastic wave
transmission problems on shells and plates \cite{NR94}. In this
context the ring frequency, that is the frequency above which
longitudinal waves in a curved shell behave as they would in a flat
plate, provides a useful point of reference.  Beyond the ring
frequency the ray dynamics in a curved shell are relatively simple
following geodesic paths, but below the ring frequency one finds
that asymptotic (ray) theories show richer features. The dispersion
relations become highly anisotropic with likewise anisotropic
propagation \cite{WHB90}. An SEA treatment would fail to capture the
non-trivial way in which the curved shell geometry influences the
wave and ray propagation below the ring frequency \cite{L94}, and
hence ray methods can provide useful insight \cite{NR94}.

Wave scattering from discontinuous line joints below the ring
frequency has been considered in \cite{L94,N98}. However,
manufacturers of large built-up mechanical structures are
increasingly developing larger and lighter sub-components, whereby
large thin shell structures are replacing more traditional
plate-beam and multi-plate assemblies. The manufacturing process for
such thin shells often entails casting molten metal (for example
aluminium), which gives rise to curved components with smooth
transitions between flat and curved regions. This raises the
question of how the ray and wave scattering, and hence the
vibrational properties of the structure, are modified in these
smooth designs. In this work we study the case of a singly curved
shell, chosen here as an assembly of two plates joined smoothly with
a quarter of a cylinder. This simplified assembly represents a
typical curved region within one of the larger sub-structures
described above. We shall go beyond plane waves and ray tracing
calculations by solving the full wave scattering problem
numerically.

The numerical solution to the full wave scattering problem will be
discussed in comparison with the corresponding ray tracing
calculations. In both cases we find effective laws for the
scattering properties, which may be inserted into ray or wave
propagation modelling techniques such as dynamical energy analysis
(DEA) \cite{GT09,C14} or the wave and finite element method
\cite{EM09, JR13}. In particular, combining these local scattering
models within a larger model of a built-up structure will lead to a
hybrid method for structures including curved thin shell components
in the mid-frequency regime. Here, the natural definition of the
mid-frequency regime is given by the range of frequencies that are
high enough for a pure FEM analysis to be impractical, but low
enough so that a simple geodesic description of the trajectory
evolution is invalid. In the high-frequency case, DEA \cite{GT09}
can be applied to model the vibrational energy transport of a
built-up structure along geodesic paths using the mesh data from a
FEM analysis \cite{C14}. In fact, DEA presents a link between ray
tracing and SEA by casting the wave or ray problem into the language
of evolution operators. We note that an equivalent operator
formalism has also been long known in computer graphics \cite{JK86},
although the theory underlying DEA arose from the more general
setting of evolution operators for transporting flows in dynamical
systems \cite{G98, Cvi12}.

The organisation of the article is as follows: we introduce the
necessary shell theory and derive a wave scattering model for a
singly curved shell in \refsec{sec2}. We then present two approaches
for solving the wave scattering model; a short wavelength asymptotic
ray tracing model based on this shell theory is detailed in
\refsec{sec:RayTrac}, and a finite difference discretisation of the
full wave model is described in \refsec{sec:ScatScheme}. We then
discuss and compare numerical results for both the wave and ray
scattering models, and the resulting reflection and transmission
laws in \refsec{sec:Numer}.

\section{Thin shell wave theory}\label{sec2}

\subsection{Governing equations of Donnell's shell theory}

The thin shell theory of Donnell is one of the simplest and widely
adopted models \cite{NR94,P91}. In this theory, moments and
transverse forces are expressed by the displacement $w$ of the
middle surface as known from the theory of laterally loaded plates.
As with other theories of continuum mechanics, shell theory is
formulated in tensor form \cite{F72}. Some properties of tensors are
summarised in \ref{sec:Tensors}. We assume an isotropic shell of
thickness $h$, Young's modulus $E$, density $\rho$ and with Poisson
ratio $\nu$. The displacement vector of a point originally on the
mid-surface of the shell is decomposed into tangential and normal
components thus $ \mathbf{u} = [u^1\: u^2\:w]^\text{T}$.

The following tensor equation for the normal displacement $w$ may be
derived \cite{NR94}
\begin{align}
\label{eq:pdeSys1} \rho h \frac{\partial^2 w}{\partial t^2} =& -
D_\alpha D_\beta(B (1-\nu) D^\alpha D^\beta w)-D_\alpha D^\alpha(B \nu D_\beta D^\beta w) \\
\nonumber & - C((1-\nu) d^\alpha_\beta \epsilon^\beta_\alpha + \nu
d^\alpha_\alpha \epsilon^\beta_\beta),
\end{align}
where \begin{equation} B=\frac{E h^3}{12(1-\nu^2)} \quad \text{ and}
\quad C= \frac{E h}{1-\nu^2} \end{equation} are the bending and
extensional stiffness, respectively. All Greek alphabet indices take
values from the set $\{1,\:2\}$. Also, the membrane strain is
\begin{equation} \label{eqn2} \epsilon_{\alpha \beta} =
\frac{1}{2}(D_\alpha u_\beta+D_\beta u_{\alpha})+ d_{\alpha \beta}
w. \end{equation} with $d_{\alpha \beta}$ the second fundamental
form and $D^\alpha$ the covariant derivative. These are discussed
further in the next section, where they are simplified for a singly
curved shell. The tangential displacements $(u^1,\:u^2)$ in the
directions $(x^1,x^2)$, respectively, satisfy \cite{NR94}
\begin{align}
\label{eq:pdeSys2} \rho h \frac{\partial^2 u^\alpha}{\partial t^2}
=& D_\beta(C((1-\nu) \epsilon^{\alpha \beta}+ \nu
\epsilon^\gamma_\gamma  g^{\alpha \beta})).
\end{align}
The term $g^{\alpha \beta}$ represents the inverse of the first
fundamental form as explained in \ref{sec:Tensors}. Note that the
above formulation is stated for a shell \emph{in vacuo} and,
although beyond the scope of the present study, coupling to an
acoustic fluid may be included via an additional term in equation
(\ref{eq:pdeSys1}) to express the pressure difference on either side
of the shell; see for example Ref. \cite{NR94}. In addition,
structural damping may be incorporated in the usual way by replacing
the Young's modulus $E$ in the equations above with
$E(1+\mathrm{i}\eta)$, where $\eta$ is the damping loss factor.
However, we proceed with the lossless case $\eta=0$ in order to
isolate the effect of curvature on the reflection and transmission
properties of the shell, whilst also allowing us to check that our
results conserve energy in the correct manner.

\subsection{Simplified model for a singly curved shell}

\begin{figure}
\centering
\begin{minipage}[!h]{0.8\linewidth}
\includegraphics[width=\textwidth]{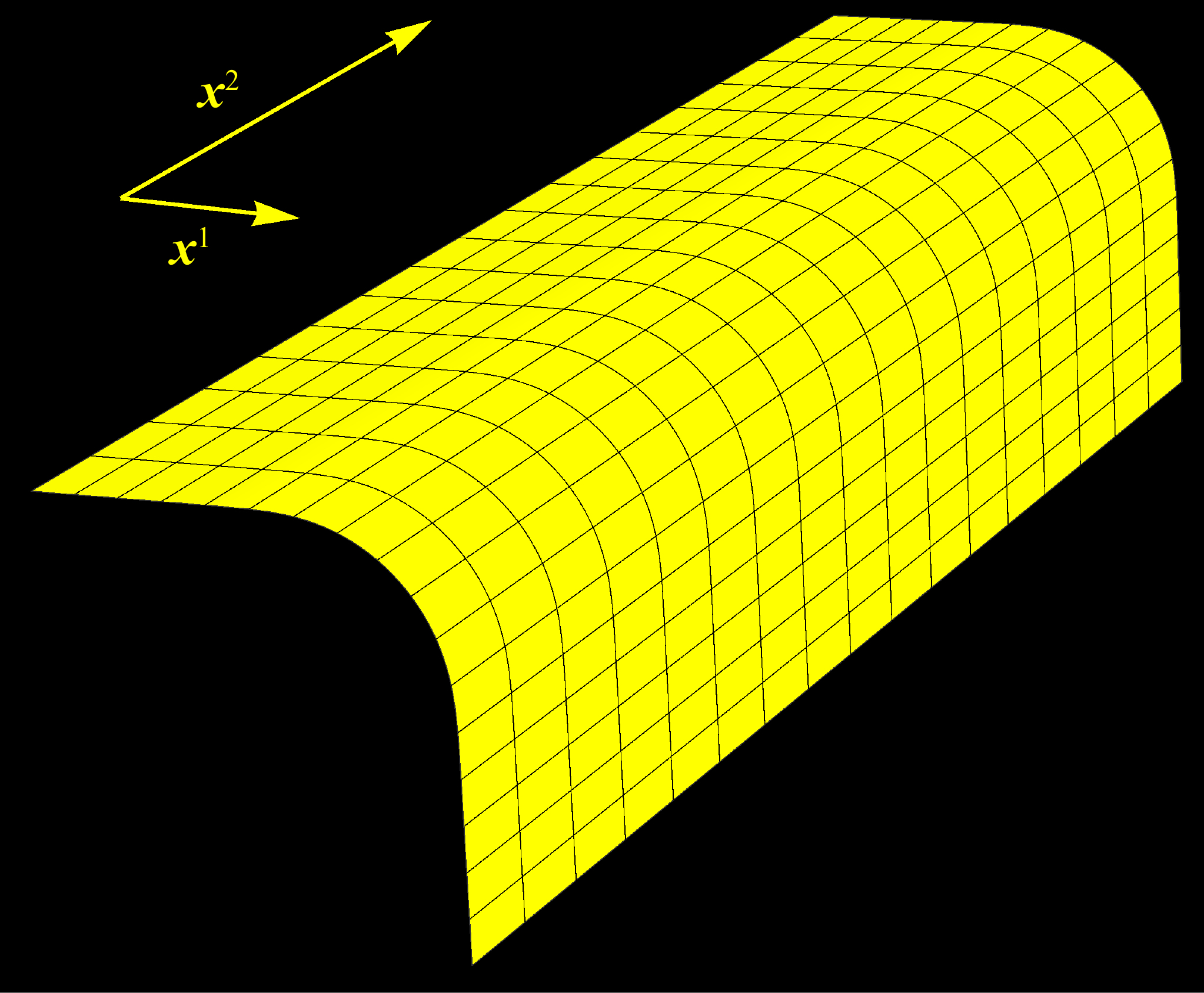}
\caption{The problem setting for which we derive a simplified set of
shell equations: a cylindrical ridge connected to flat plates on
either side.} \label{fig:ridgeGeom}
\end{minipage}
\end{figure}

\begin{figure}
\centering
\begin{minipage}[!h]{0.8\linewidth}
\includegraphics[width=\textwidth]{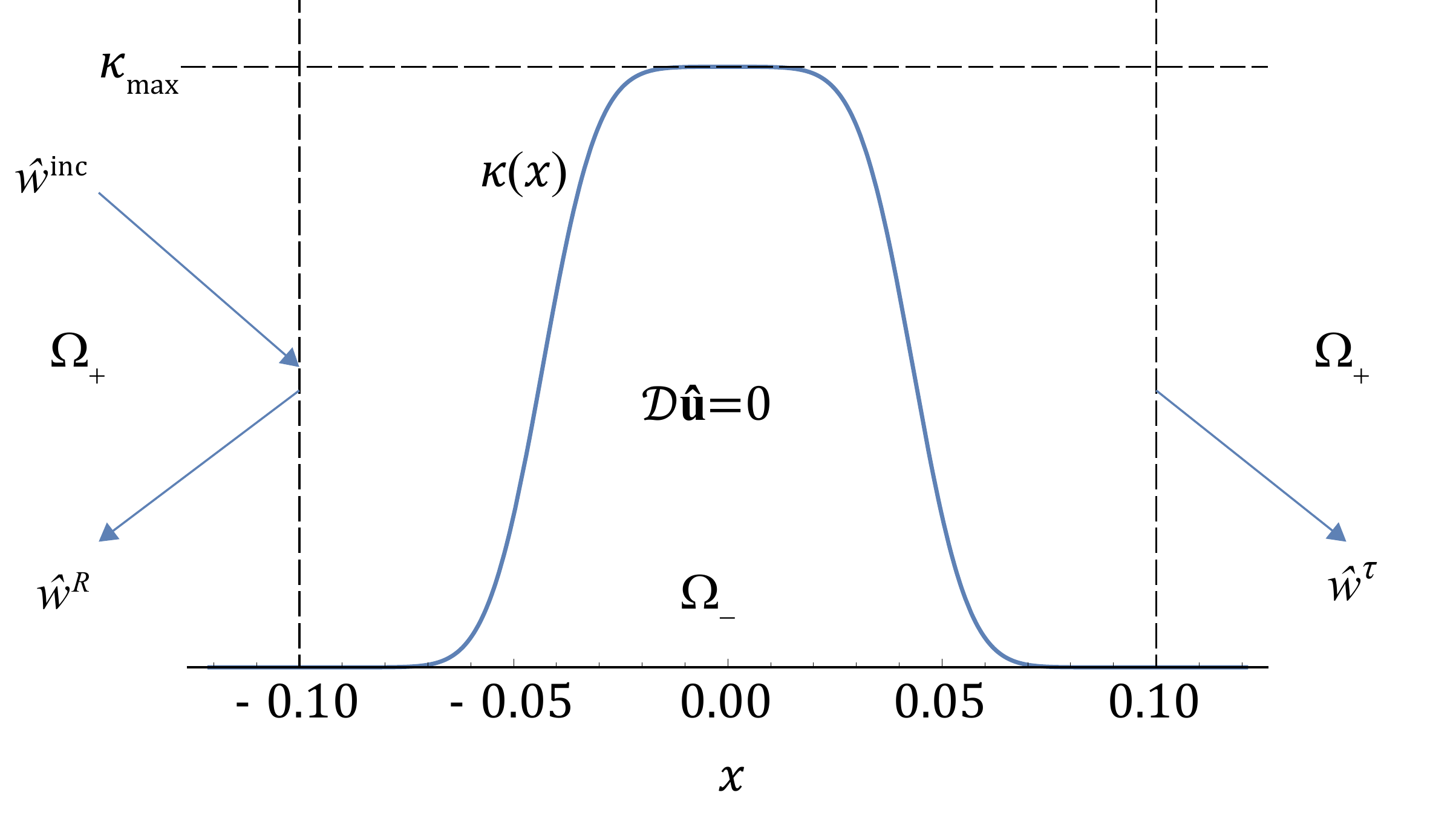}
\caption{The subdivision of the cylindrical ridge geometry into
interior ($\Omega_-$) and exterior ($\Omega_+$) regions for the
scattering problem. The interfaces between $\Omega_-$ and $\Omega_+$
lie with the flat regions where $\kappa(x)\approx 0$. For
$x\in\Omega_-$, the curvature increases smoothly to
$\kappa(x)=\kappa_{\mathrm{max}}$ at $x=0$. The incident, reflected
and transmitted waves in $\Omega_+$ are also indicated.}
\label{fig:Curv}
\end{minipage}
\end{figure}

In this Section we simplify the shell theory presented above for the
case of a cylindrical ridge as shown in Fig. \ref{fig:ridgeGeom}.
The only simplifying assumption is that the shell is not curved with
respect to the direction $x^2$. We define the principal curvature
$\kappa_1$ (in the direction $x^1$) as
\begin{equation}\label{eq:curv1}
\kappa_1(x^1)=\frac{f(x^1)}{f(0)} \kappa_{\mathrm{max}},
\end{equation}
where $\kappa_{\mathrm{max}}$ is a constant corresponding to the
maximum curvature in the cylindrical region. A smoothly varying
curvature with respect to $x^1$ is obtained via the interpolation
function $f$ given by
\begin{equation}\label{eq:interp}
f(x^1)= \frac{1}{2} \left( \mathrm{erf} \left(\frac{x^1+x^*}{\delta
x} \right)- \mathrm{erf}\left(\frac{x^1-x^*}{\delta x}\right)
\right),
\end{equation}
where $x^1=\pm\: x^*$ are the centres of the transitions between the
two flat regions where $\kappa_1=0$ and the cylindrical region where
$\kappa_1=\kappa_{\mathrm{max}}$, see \reffig{fig:Curv}. Also,
$\delta x$ is the width of the transition region. Note that the case
of a non-smooth joint between a cylindrical ridge and a pair of
connecting plates is obtained in the limit $\delta x\rightarrow0$.

A cylindrical ridge with a prescribed bending angle $\psi$ can be
designed by integrating the angular increment
\begin{equation}
\mathrm{d}\psi = \kappa_1 \mathrm{d}x^1
\end{equation}
to find the total length $L$ of the curved section of the geometry
such that
\begin{equation} \label{eq:angleInc}
\psi = \int_{-{L/2}}^{{L/2}}\kappa_1 \mathrm{d}x^1
\end{equation}
for the desired angle. The length $L$ is then found by numerically
solving the implicit equation \refeq{eq:angleInc}. In this work we
take a quarter cylinder with $\psi=\pi/2$ as shown in Fig.
\ref{fig:ridgeGeom}.

In the above described geometrical setting, the shell theory
presented in the last section is simplified considerably if we
choose an orthonormal basis $(\mathbf{a}_1,\mathbf{a}_2)$ along the
axes $x^1,x^2$. We then measure all tensor fields relative to this
basis, that is, we work in physical components. The metric becomes
\begin{equation}
\mathrm{d}s^2 = (\mathrm{d}x^1)^2+(\mathrm{d}x^2)^2
\end{equation}
and so the first fundamental form and its dual both reduce to the
identity matrix \begin{equation} \label{eq:MetricId} g_{\alpha
\beta} := \mathbf{a}_\alpha \cdot \mathbf{a}_\beta= \delta_{\alpha
\beta} = g^{\alpha \beta}.\end{equation} Therefore the raising and
lowering of indices becomes trivial. The second fundamental form
$d_{\alpha \beta}$ in the basis $(\mathbf{a}_1,\mathbf{a}_2)$ is
simply \begin{equation} d_{1 1} = \kappa_1, \end{equation} and zero
for all other combinations of $\alpha$ and $\beta$.

The covariant derivative (see \ref{sec:Tensors}), likewise reduces
considerably since all derivatives vanish for the metric
\refeq{eq:MetricId}. Hence, all Christoffel symbols vanish and the
covariant derivative is identical to the directional derivative,
i.e. $D_\alpha =\partial_\alpha$. Applying the above simplifications
in the shell equation for the normal displacement (\ref{eq:pdeSys1})
and assuming that material constants and the thickness are constant
yields the following equation
\begin{align}\label{eq:pdeSys1a}
\rho h \frac{\partial^2 w}{\partial t^2} = - B\triangle^2w
 - C \kappa\left(\frac{\partial u^1}{\partial x}+\nu\frac{\partial u^2}{\partial
 y}+\kappa w\right),
\end{align}
where we write $(x^1,x^2)\equiv(x,y)$ and
$\kappa_1(x^1)\equiv\kappa(x)$ for simplicity of notation. We adopt
this notation for the remainder of the paper since we are no longer
considering differential equations written in tensor form. Repeating
the simplification process for the in-plane wave equation system
(\ref{eq:pdeSys2}) gives
\begin{align}
\label{eq:pdeSys2a} \frac{\rho h}{C} \frac{\partial^2 u^1}{\partial
t^2} =&\:\frac{\partial^2 u^1}{\partial x^2}+\frac{(1+\nu)}{2}\frac{\partial^2u^2}{\partial x \partial y}+\frac{(1-\nu)}{2}\frac{\partial^2u^1}{\partial y^2}+\frac{\partial}{\partial x}(\kappa w),\\[3pt]
\label{eq:pdeSys2b}\frac{\rho h}{C}  \frac{\partial^2 u^2}{\partial
t^2} =& \:\frac{\partial^2 u^2}{\partial
y^2}+\frac{(1+\nu)}{2}\frac{\partial^2u^1}{\partial x
\partial y}+\frac{(1-\nu)}{2}\frac{\partial^2u^2}{\partial
x^2}+\nu\kappa\frac{\partial w}{\partial y}.
\end{align}
Hence, a simplified set of partial differential equations (PDEs)
describe the wave motion in the cylindrical ridge shell geometry
under consideration here. In the next section we discuss how under
certain modelling assumptions, this system can be reduced further to
a set of ordinary differential equations (ODEs).

\subsection{Reduction to a system of ordinary differential equations}

In this work we are primarily interested in how the scattering
properties of a thin shell cylindrical ridge depend on the frequency
and the direction of a prescribed incident plane wave originating in
the flat region (where $\kappa\sim0$). As such it makes sense to
consider time-harmonic waves with angular frequency $\omega$.
Further, we can extract scattering properties that are independent
of the position along the ridge by exploiting the translational
symmetry and assuming that the ridge is of infinite extent in the
$y-$direction. Under these assumptions we may write
\begin{equation}\label{eq:ODEansatz}
\mathbf{u}(x,y,t) = \mathbf{\hat{u}}(x) e^{\mathrm{i} (k_y y -
\omega t)} \,,
\end{equation}
where $k_y$ is the component of the wavenumber in the $y-$direction
and $\mathbf{\hat{u}}=[u\hspace{3pt}v\hspace{3pt}\hat{w}]^\text{T}$ are the
coefficients of (\ref{eq:ODEansatz}) in the in-plane directions $x$
and $y$, and the normal direction, respectively. Substitution of the
ansatz (\ref{eq:ODEansatz}) into the PDE system (\ref{eq:pdeSys1a})
- (\ref{eq:pdeSys2b}) yields the following fourth-order ODE system
in the variable $x$:
\begin{align}\label{eq:odeSys1}
c_p^2\:\frac{\ud^2 u}{\ud
x^2}+\left(\omega^2-c_s^2k_y^2\right)u+\mathrm{i}
k_y(c_p^2-c_s^2)\frac{\ud v}{\ud x}+c_p^2\frac{\ud}{\ud x}(\kappa \hat{w})&=0,\\[2pt]\label{eq:odeSys2}
c_s^2\:\frac{\ud^2 v}{\ud
x^2}+\left(\omega^2-c_p^2k_y^2\right)v+\mathrm{i}
k_y\left((c_p^2-c_s^2)\frac{\ud
u}{\ud x}+\nu c_p^2\kappa \hat{w}\right)&=0,\\[2pt]\label{eq:odeSys3}
\frac{B}{\rho h}\left(\frac{\ud^4 \hat{w}}{\ud
x^4}-2k_y^2\frac{\ud^2 \hat{w}}{\ud
x^2}+k_y^4\hat{w}\right)+(c_p^2\kappa^2-\omega^2)\hat{w}+c_p^2\kappa\left(\frac{\ud
u}{\ud x}+i\nu k_y v\right)&=0.
\end{align}
Note that the constants in the above system have been simplified by
writing them in terms of the pressure and shear wave velocities,
$$c_p=\sqrt{\frac{E}{\rho(1-\nu^2)}} \hspace{5mm}\text{and}\hspace{5mm}
c_s=\sqrt{\frac{E}{2\rho(1+\nu)}},$$ respectively. In particular, we
have made use of the following easily verified relations
$$c_p^2=\frac{C}{\rho h},\hspace{5mm}\frac{c_s^2}{c_p^2}=\frac{(1-\nu)}{2}\hspace{5mm}\text{and}\hspace{5mm} c_p^2-c_s^2=\frac{(1+\nu)c_p^2}{2}.$$
Later in the paper we will write the above ODE system in the
shorthand form \begin{equation}\label{eq:ODEsysB} \mathcal{D}
\mathbf{\hat{u}}=0, \end{equation} for brevity of exposition.

\subsection{Formulation of the scattering problem} \label{sec:Scat}

We connect the curved region to flat plates on each side using
interfaces. Imposing a set of conditions at these interfaces will
enable us to formulate the set of ODEs (\ref{eq:ODEsysB}) as a
boundary value problem, and then by considering the flux of the
incoming and outgoing wave fields at these interfaces we are able to
formulate a scattering problem. We assume that the interfaces reside
in the asymptotically flat regions and as such the interfaces
themselves do not give rise to reflection/transmission phenomena,
only the interior region between the interfaces governs the
scattering properties.

Each interface is assumed to satisfy the  continuity conditions
given in Ref. \cite{N98}, that is, continuity of displacement,
rotation, traction, moment, and normal shear stress. Since the
material properties are constant throughout the entire geometry,
then the interface conditions may be written simply as
\begin{align}\label{eq:IF1}
  \mathbf{\hat{u}}^- &= \mathbf{\hat{u}}^+, \\\label{eq:IF2}
\frac{\ud \mathbf{\hat{u}}}{\ud x}^- &= \frac{\ud
\mathbf{\hat{u}}}{\ud x}^+, &
\\\label{eq:IF3}
 \frac{\ud^2 \hat{w}}{\ud x^2}^- &= \frac{\ud^2 \hat{w}}{\ud x^2}^+, &\\\label{eq:IF4}
 \frac{\ud^3 \hat{w}}{\ud x^3}^- &= \frac{\ud^3 \hat{w}}{\ud x^3}^+. &
\end{align}

The superscripts specify the value of the quantity as we approach
the interface from either the interior region $(-)$ containing the
ridge, or the exterior flat region $(+)$ beyond the interfaces on
either side of the ridge. In the sequel we will refer to the
interior region as $\Omega_{-}$, and the union of the exterior
regions as $\Omega_+$ (see \reffig{fig:Curv}).

The waves in $\Omega_+$ are precisely those of classical plate
theory and therefore the wave modes in $\Omega_+$ that are scattered
by $\Omega_-$ will be one (or more) of bending, pressure, shear or
evanescent bending type. In relation to the vector
$\mathbf{\hat{u}}$ we have that $\hat{w}$ describes the sum of the
bending wave contributions and that the in-plane wave types will
each be given by a linear combination of $u$ and $v$. In this work
we consider only incident bending modes $\hat{w}^{\text{inc}}$
originating in the exterior $\Omega_+$ and being scattered by the
ridge in $\Omega_{-}$. However, the extension to other incident wave
types is straightforward. For the purpose of investigating
directional properties we consider plane wave scattering. We
introduce the notation that the interface to the left of
$\Omega_{-}$ is located at $x=x^l$ and the interface to the right is
correspondingly at $x=x^r$. Then an incident wave of unit amplitude
travelling in from the left of $\Omega_-$ can be written in the form
\begin{align}\label{eq:winc}
\hat{w}^{\text{inc}}(x)=\exp(\mathrm{i} k^b_x (x-x^l)), \,
\end{align}
where $k^b_x$ is the wave-number associated with the incident
bending mode in the $x$-direction. In order to write the resulting
scattered waves in a concise manner we introduce $x_\tau=x-x^r$ and
$x_R=x-x^l$. Then the scattered waves in $\Omega_+$ may be written
\begin{align}
\varphi_{\beta}^{\alpha}(x)=A^\alpha_\beta\exp(\pm \mathrm{i}
k^\alpha_x x_\beta), \,
\end{align}
where $\alpha\in\{b, e, p, s \}$ designates the scattered wave type
as either bending ($b$), evanescent bending ($e$), pressure ($p$) or
shear ($s$). The symbol $\beta\in\{R,\tau\}$ designates whether the
scattered wave is reflected ($R$) or transmitted ($\tau$). That is,
whether the scattered wave emerges in $\Omega_+$ on the same  side
of the ridge as the incident wave is sent in, or on the other side.
The value of $\beta$ also prescribes the sign in the $\pm$ as
negative for $\beta=R$ and positive for $\beta=\tau$. The
coefficient $A^\alpha_\beta$ denotes the corresponding wave
amplitude, which are related to $\mathbf{\hat{u}}$ via the following
relations
\begin{align}
u^{\beta}(x)=& \pm A^p_\beta\exp(\pm \mathrm{i} k^p_x
x_\beta)\cos\theta-A^s_\beta\exp(\pm \mathrm{i} k^s_x x_\beta) \sin\theta,\\[2pt]
v^{\beta}(x)=&A^p_\beta\exp(\pm \mathrm{i} k^p_x
x_\beta)\sin\theta\pm A^s_\beta\exp(\pm \mathrm{i} k^s_x
x_\beta) \cos\theta,\\[2pt]\label{eq:wRT}
\hat{w}^{\beta}(x)=&A^b_\beta\exp(\pm \mathrm{i} k^b_x
x_\beta)+A^e_\beta\exp(-k^b_x |x_\beta|).
\end{align}
Here $\theta\in(-\pi/2,\pi/2)$ is the angle between the scattered
wave directions and the $x-$axis (positive or negatively oriented
depending on the direction of propagation). Hence, the total wave
field in $\Omega_+$ to the left of $\Omega_-$ is given by
\begin{equation}
\mathbf{\hat{u}}^l=[u^{R}\hspace{5pt} v^{R}\hspace{5pt}
\hat{w}^{R}+\hat{w}^{\text{inc}}]^\text{T},
\end{equation}
and the total wave field to the right of $\Omega_-$ may be written
\begin{equation}
\mathbf{\hat{u}}^r=[u^{\tau}\hspace{5pt} v^{\tau}\hspace{5pt}
\hat{w}^{\tau}]^\text{T}.
\end{equation}

The scattering problem is then formed by connecting the interior
problem for $\mathbf{\hat{u}}=\mathbf{\hat{u}}^-$ in $\Omega_-$ with
the plane wave ansatz in the exterior regions for
$\mathbf{\hat{u}}^l$ and $\mathbf{\hat{u}}^r$ described above, via
the interface conditions (\ref{eq:IF1}) to (\ref{eq:IF4}). Here
$\mathbf{\hat{u}}^l$ and $\mathbf{\hat{u}}^r$ take the role of
$\mathbf{\hat{u}}^+$ on the left and right interfaces, respectively.
When this scattering problem is solved, we can extract the
scattering solutions $\mathbf{\hat{u}}^l$ and $\mathbf{\hat{u}}^r$
and the scattering coefficients $A^\alpha_\beta$, the latter of
which are of primary interest for this study.

\section{From waves to rays: short wavelength asymptotics} \label{sec:RayTrac}

One obtains a ray tracing model from the PDE model
(\ref{eq:pdeSys1a}) - (\ref{eq:pdeSys2b}) by moving to short
wavelength asymptotics using the ansatz
$$\mathbf{u}(x,y,t)=\mathbf{u}^{(\varepsilon)}(x,y,t)\exp\left(\mathrm{i}\varepsilon^{-\lambda}\phi(x,y,\varepsilon^{\mu} t)\right),$$
where $\phi$ is a phase function and $\varepsilon$ is a small
parameter. The choice of the parameters $\lambda\geq0$ and
$0\leq\mu\leq\lambda$ determines the wave type, bending or in-plane.
Then we define the frequency
$\omega=-\partial_t\phi(x,y,\varepsilon^{\mu} t)$ and the wavenumber
vector $\mathbf k = \nabla\phi$. Note that the pre-factors of $t$
vanish after applying the asymptotic scaling and setting
$\lambda=\mu=0$. Assuming that $k=|\mathbf{k}|$ is large in
comparison to the curvature, Pierce derived a general dispersion
relation \cite{P91} that was later presented in a simpler form by
Norris and Rebinsky \cite{NR94}. In the next section we apply this
dispersion relation to generate the Hamiltonian dynamics for our ray
tracing model on a singly curved shell.

\subsection{Below the ring frequency}

The ring frequency for the cylinder is defined relative to the
pressure mode and corresponds to the frequency above which a
longitudinal wave can traverse around the cylinder. This extensional
motion gives rise to breathing modes in cylinders, which also result
in a radial motion. In the vicinity of the ring frequency,
approximate forms of the dispersion relations are obtained for the
in-plane and bending modes using different scalings. These
dispersion relations each reduce to those for a flat plate when the
frequency sufficiently exceeds the ring frequency. Hence in this
study, we refer to the mid-frequency case as the the frequency range
in the vicinity of the ring frequency and the high frequency regime
as the frequencies that are sufficiently large for the dispersion
relations to reduce to those of a flat plate. Defining the
longitudinal wavenumber to be ${\mit\Omega}=\omega/c_p$, with $c_p$ the
longitudinal plate wave speed as before, and denoting the
corresponding wavenumber at the ring frequency as ${\mit\Omega}_{*}$, we
find that ${\mit\Omega}_{*} = \kappa$ for the singly curved shell
configuration shown in Fig.\ \ref{fig:ridgeGeom}.

\begin{figure}
\centering
\includegraphics[width=\textwidth]{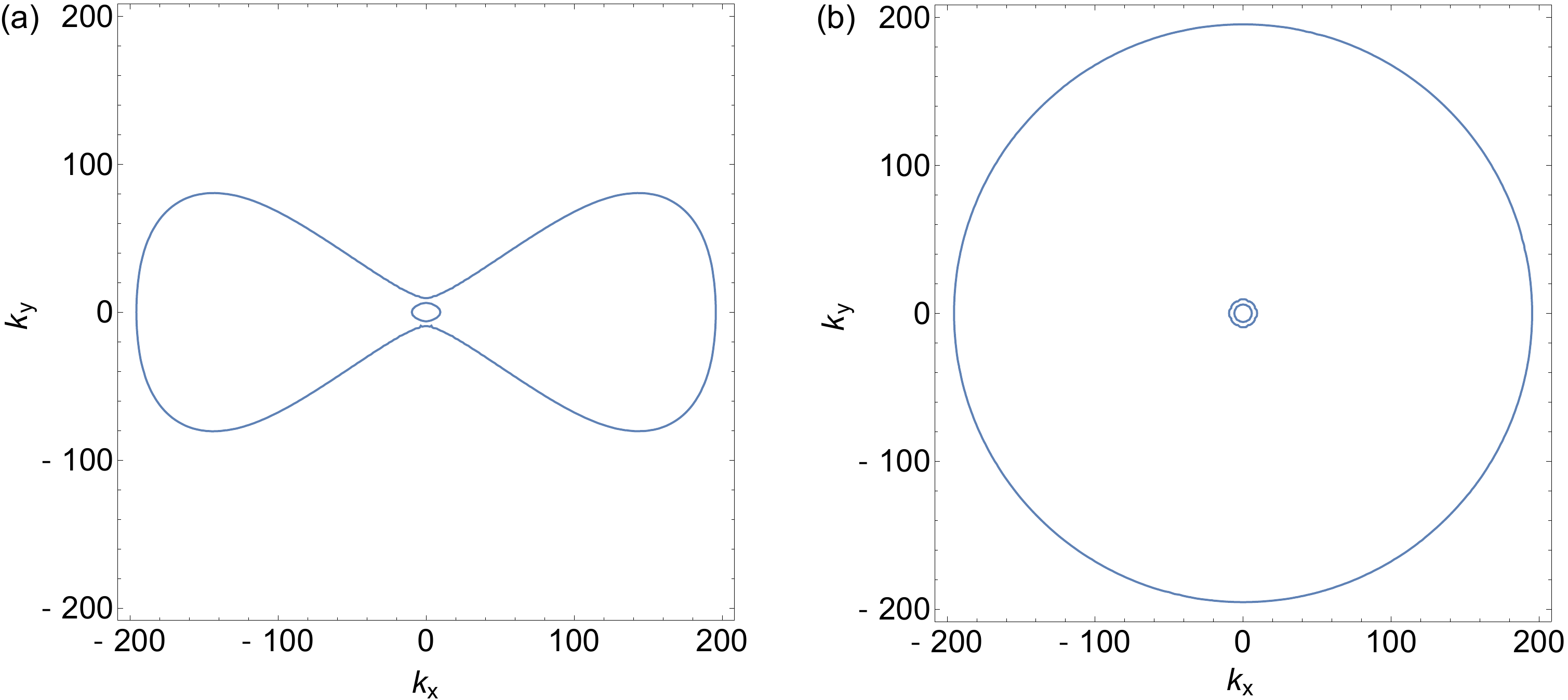}
\caption{Dispersion relations represented by plotting
${\mit\Omega}(k_x,k_y)$ at zeros of the Hamiltonian $\tilde{H}$. (a)
Anisotropic dispersion relation for a cylinder. (b) Isotropic
dispersion relation for a flat plate.}\label{fig:disp}
\end{figure}

Below the ring frequency, the full dispersion relationship is
usually considered. For the configuration in Fig.\
\ref{fig:ridgeGeom} with zero curvature in the $y-$ direction we
obtain the following expression from Ref. \cite{NR94} in physical
coordinates:
\begin{align} \label{Ham}
\tilde{H}(x,k, \omega)=&\left( {\mit\Omega}^2 - \frac{1}{2} k^2 (1-\nu)
\right)\left( ({\mit\Omega}^2-k^2) \left({\mit\Omega}^2 - \frac{h^2 k^4}{12}
\right)- {\mit\Omega}^2
{\mit\Omega}_{*}^2\right)\\
 \nonumber & + (1-\nu^2) \left( {\mit\Omega}^2\kappa^2k_y^2
-\frac{1}{2} (1- \nu)\kappa^2 k_y^4 \right) \, .
\end{align}
Two extreme cases of the dispersion curve are depicted in Fig.\
\ref{fig:disp}, which shows the respective dispersion relations for
a cylinder and for a flat plate using the parameters given in
\ref{sec:Parameters}. The outer part of the cylinder dispersion
curve has also been observed experimentally \cite{WHB90}.

\subsection{Hamiltonian system}
The Hamiltonian $\tilde{H}$ (\ref{Ham}) gives rise to the following
ODE system, which governs the ray dynamics on a singly curved shell
\begin{align}
\dot{x} =&  \phantom{aa} \frac{\partial \tilde{H}}{\partial k_x}, \\
\dot{y} =&  \phantom{aa} \frac{\partial \tilde{H}}{\partial k_y}, \\
\dot{k}_x =& -\frac{\partial \tilde{H}}{\partial x},\\
\dot{k}_y =& -\frac{\partial \tilde{H}}{\partial y}\,.
\end{align}
In a full time-domain simulation one would also have two additional
equations
\begin{align}\label{eq:Hex1}
\dot{t}  \phantom{a} =&-\frac{\partial \tilde{H}}{\partial \omega},
\\\label{eq:Hex2} \dot{\omega}  =& \phantom{aa}\frac{\partial
\tilde{H}}{\partial t} \,.
\end{align}
In this work however, the frequency is constant due to the fact our
Hamiltonian $\tilde{H}$ is time-independent. Furthermore, since we
also take the material parameters to be constants throughout the
shell, equation (\ref{eq:Hex1}) simply represents a
re-parametrisation of time to the fictitious time used here.
Equations (\ref{eq:Hex1}) and (\ref{eq:Hex2}) are therefore not
needed in the time-harmonic description. Finally, due to the
translational invariance in the $y$-direction, $k_y$ is constant and
it suffices to study the above Hamiltonian system in the $(x,k_x)$
phase-plane only.

\section{Discretisation of the wave scattering problem}\label{sec:ScatScheme}

In order to solve the full wave scattering problem we discretise the
ODE system in $\Omega_-$ using finite difference methods and couple
this to the scattering of incoming and outgoing waves in $\Omega_+$
using the interface coupling conditions as described below.

\subsection{Finite difference method in $\Omega_-$}

The differential operator $\mathcal{D}$ in the ODE system
(\ref{eq:ODEsysB}) includes both bending and in-plane waves, along
with the coupling between them. Each of the equations in the system
is discretised using second order accurate centered finite
difference (FD) formulae on a set of equi-spaced grid points
$\{x_j\}$ with $j=0,\dots,N$. We denote the centered finite
difference matrices for approximating derivatives of order $n$ by
$\mathbf{D}_0^n$ and define
\begin{equation} \opkappa = [\kappa(x_i)
\delta_{i j}]\hspace{5mm} \text{ with }\:\: i,j=0,\dots,N \,
\end{equation}
to be a diagonal matrix containing the curvature values.

To create a vectorial ODE discretisation we use the Kronecker
product $\otimes$, see for example Ref. \cite{CL00}, along with the
following $3 \times 3$ projector matrices:
\begin{align}
\opE_{i j} &= [\delta_{i a} \delta_{j b}] \:\text{ for }\: a,b=1,2,3, \\
\opPi_i &= \opE_{i i}, \\
\opH_{2 3} &= \left[ \begin{array}{ccc}  0 & 0 & 0 \\ 0& 0 & \mathrm{i}\\ 0
& - \mathrm{i}& 0 \end{array} \right] \,.
\end{align}
Using the above notation we can write the various operators of our
ODE in a compact way. The in-plane equations (\ref{eq:odeSys1}) and
(\ref{eq:odeSys2}) will have a diagonal (self-interaction) and
frequency independent part
\begin{equation}
\mathbf{M}_{\text{diag}}=(c_p^2 \opD_0^2 -c_s^2k_y^2 \mathbf{I})
\otimes \opPi_1 + (c_s^2 \opD_0^2-c_p^2k_y^2 \mathbf{I} ) \otimes
\opPi_2,
\end{equation}
where $\mathbf{I}$ is the $3 \times 3$ identity matrix. The coupling
between the $u$ and $v$ components is symmetric and is given by
\begin{equation}
\mathbf{M}_{\text{off}} = \mathrm{i} k_y (c_p^2-c_s^2) \opD_0^1
\otimes\left(\opE_{12}+\opE_{21}\right).
\end{equation}
The bending wave equation (\ref{eq:odeSys3}) has a diagonal
(self-interaction) and frequency independent part
\begin{equation}
{\mathbf{B}}=-\left(\frac{B}{\rho h}(\opD_0^4 - 2 k_y^2 \opD_0^2 +
k_y^4 \mathbf{I})+ c_p^2 \opkappa^2  \right) \otimes  \opPi_3 \,,
\end{equation}
and the coupling between the bending and in-plane components is
given by
\begin{equation}
\mathbf{C}= c_p^2 ( \opD_0^1 \opkappa \otimes \opE_{1 3} -\opkappa
\opD_0^1 \otimes \opE_{3 1}+k_y \nu \opkappa \otimes \opH_{2 3})\,.
\end{equation}
The  shell operator $\mathcal{D}$ then has the following discrete
representation:
\begin{equation}
\mathbf{D} =\mathbf{M}_{\mathrm{diag}}+
\mathbf{M}_{\mathrm{off}}+\mathbf{B}+\mathbf{C} + \omega^2
\mathbf{I} \, .
\end{equation}

The FD operators will extend beyond the grid over $\Omega_-$ as the
stencil reaches into the scattering region. This leads to a coupling
between the degrees of freedom in the finite difference
approximation and the unknown scattering amplitudes
$A^\alpha_\beta$. Conversely, the incident wave in $\Omega_+$ will
produce a forcing term which drives the finite difference
calculation in $\Omega_-$. This coupling of the FD solution in
$\Omega_-$ and the plane wave description in $\Omega_+$ is discussed
further below.

\subsection{Coupling of the interior and exterior regions}

The scattering solution and its derivatives in $\Omega_+$ are
matched to the FD solution in $\Omega_-$ at the interfaces between
$\Omega_-$ and $\Omega_+$ according to the conditions (\ref{eq:IF1})
to (\ref{eq:IF4}). The incident wave gives rise to inhomogeneous
terms for the excitation of the FD model at the left interface. The
derivatives appearing in the coupling conditions are implemented in
the FD-solution in $\Omega_-$ using one-sided finite difference
operators with second order accuracy. Forward difference formulae
are employed at the left interface and backward difference formulae
are employed at the right interface so that the stencils extend only
into $\Omega_-$. The extremal points of these stencils are taken at
the scattering boundaries, where the value of the scattering
solution is applied as a boundary condition. The interfaces are thus
taken to be at positions $\Delta x$ away from the extremal inner
grid points, where $\Delta x$ is the discretisation step in the FD
approximation.

As an illustrative example we consider the condition on
$\mathrm{d}^3\hat{w}/\mathrm{d}x^3$ \refeq{eq:IF4} at the interface
receiving the incident wave (\ref{eq:winc}). The reflected bending
modes include both a propagative contribution $A^b_R\exp\!\left(-\mathrm{i}
k_x^{b} x_R \right)$ and an evanescent contribution
$A^e_R\exp\!{\left(-k_x^{b} |x_R|\right)}$ as shown in equation
(\ref{eq:wRT}). The forward difference formula for the third
derivative takes the form \begin{equation} V(f):= D^3_+ f_0 =
\frac{1}{\Delta x^3} \left(-\frac{5}{2}f_0+9f_1-12f_2+7f_3
-\frac{3}{2}f_4\right).
\end{equation} Matching the scattering and finite difference
solutions using the interface condition (\ref{eq:IF4}) leads to
\begin{align}\label{eq:IFdisc}
\frac{\mathrm d^3}{\mathrm d x^3}\left(\hat{w}^{\text{inc}} +
A^b_R\exp\!\left(-\mathrm{i} k_x^{b} x_R \right)
+A^e_R\exp\!{\left(- k_x^{b} |x_R|\right)}\right) &= V(\hat{w}).
\end{align}
Splitting the discretised boundary operator $V$ into a part that
acts only on the left-most entry at the interface and one that acts
on the remaining nodes in $\Omega_-$, we obtain \begin{equation} V =
\delta_{\mathrm{i}0}V + (1-\delta_{\mathrm{i}0})V := V_0 + \vr, \quad  i=0,\dots,4.
\end{equation} At $x=x_l=x_0$, the left hand side of (\ref{eq:IFdisc})
evaluates to
\begin{align}
-(k_x^b)^3(\mathrm{i} +\mathrm{i} A^b_R + A^e_R),
\end{align}
and using the splitting described above, the right hand side may be
written as
\begin{align}
V_0\left(1 + A^b_R +A^e_R\right) + \vr(\hat{w}).
\end{align}
Writing $\hat{w}_i\approx\hat{w}(x_i)$ for $i=1,2,3,4$ and using the
definition of $V$ yields a linear equation
\begin{equation}\label{eq:Vcond}
(k_x^b)^3(\mathrm{i} +\mathrm{i} A^b_R + A^e_R)=\frac{5(1 + A^b_R
+A^e_R)-18\hat{w}_1+24\hat{w}_2-14\hat{w}_3+3\hat{w}_4}{2\Delta x^3}
\end{equation}
in terms of the unknowns $A^b_R$,  $A^e_R$ and $\hat{w}_i$,
$i=1,2,3,4$. Continuing for the remaining boundary conditions at the
left interface $x=x_0$ yields similar relations, all following the
same pattern as in \refeq{eq:Vcond}. The boundary conditions at the
right interface give similar results for the transmitted terms, but
without the incident wave terms. Finally, we combine the discretised
interface coupling conditions with the interior finite difference
equations. This leads to the matrix problem: \begin{equation}
\left[\begin{matrix}
* & * & \mathbf{0}\\
* & \mathbf{D} & *\\
\mathbf{0}& * & *
\end{matrix}
\right] \left[\begin{matrix}
\mathbf{R}\\
\hat{\mathbf{u}}_\Delta\\
\mathbf{T}
\end{matrix}
\right]= \left[\begin{matrix}
*\\
\mathbf{0}\\
\mathbf{0}
\end{matrix}
\right] \end{equation} with scattering coefficients
\begin{equation}
\mathbf{R}=[A_R^b\:\: A_R^e\:\: A_R^p\:\: A_R^s]^\text{T},
\end{equation}
and
\begin{equation}
\mathbf{T}=[A_\tau^b\:\: A_\tau^e\:\: A_\tau^p\:\: A_\tau^s]^\text{T}.
\end{equation}
Here, the finite difference solution in the interior of $\Omega_-$
is represented by \begin{equation} \hat{\mathbf{u}}_\Delta=[u_i\:\:
v_i\:\: \hat{w}_i]^\text{T},\:\: i=1,\dots, N-1.\end{equation} Note that
the stencil for the second derivative FD operator within
$\mathbf{D}$ takes boundary data from the plane wave terms at the
interface points $x_0$ and $x_N$, whereas the fourth derivative
operator extends beyond this and also takes boundary data from
within the scattering region at $x=x_0-\Delta x$ and $x=x_N+\Delta
x$.

\section{Numerical results}\label{sec:Numer}
In this section we discuss the numerical solution of the full wave
scattering problem derived in Sect. \ref{sec2} and compare the
results to those obtained using an ODE time-stepper for the
Hamiltonian system presented in Sect. \ref{sec:RayTrac}. The
parameters chosen for all calculations are summarised in
\ref{sec:Parameters}. For the cylindrical region, these parameters
correspond to a circularly cylindrical steel shell as considered in
Refs. \cite{WHB90,RN96}. Note that whilst the width of the
cylindrical ridge studied here is on the millimeter scale, it only
represents a small region of the larger structures and components
that serve to motivate this study.

An incident bending mode in the left part of $\Omega_+$ is used to
excite the system and is sent in to the interface with $\Omega_-$ at
various angles, corresponding to a variation in the trace wave
number $k^b_y$. The latter is chosen according to the values given
by the dispersion curve. In all cases we set $\omega=9742\pi$ Rad/s,
which is large enough so that waves will always transmit straight
through the cylindrical region (as though it were a flat plate) when
$k_y^b=0$, that is when waves approach the cylindrical region
directed parallel to the $x$-axis. We note that in the high
frequency regime this behaviour will be preserved for almost all
incident waves when both $k_x^b$ and $k_y^b$ are positive. However,
the choice of $\omega$ here corresponds to the more interesting
mid-frequency case where reflections are also possible for a range
of sufficiently large $k_y^b>0$.

\subsection{Ray tracing calculations}

\begin{figure}[!h]
\centering
\includegraphics[width=\textwidth]{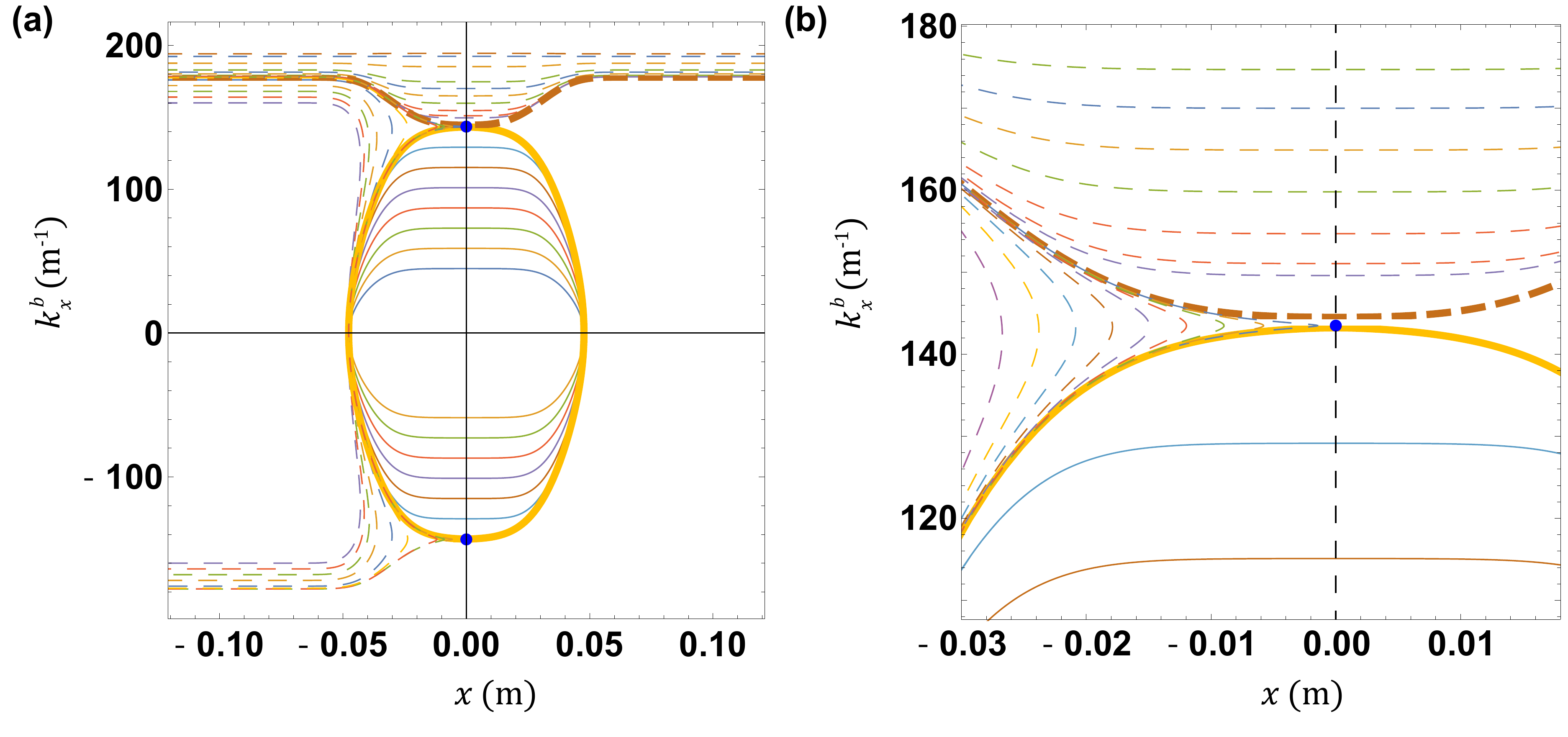}
\caption{(a) Ray trajectories represented in the $(x,k_x^b)$
phase-plane showing two fixed points at $x=0$ and the separatrix (solid bold curve) which connects them as it traverses the cylindrical ridge. (b) A close up view of the upper fixed point and separatrix region. The
upper fixed point divides the incoming rays from the left hand side (dashed lines) into reflecting or transmitting trajectories. The bold dashed line depicts the extremal transmitting trajectory closest to the set of reflecting rays. In both subplots, the solid line trajectories are those which are trapped inside the ridge.}\label{fig:PhaseSpace4}
\end{figure}

In this section we consider the reflection/ transmission behaviour
of rays corresponding to an incident bending mode (see
\reffig{fig:PhaseSpace4}). A range of incoming wavenumbers
(directions) are used corresponding to the strip in the upper left
corner of the figure. We find that the rays transmit for
sufficiently large $k^b_x$, whereas for smaller positive values of
$k^b_x$ the rays reflect. Note that the symmetry of the problem
means that rays also transmit for $k^b_x$ negative when the incoming
ray is from the right hand side and $|k^b_x|$ is large enough. The
threshold value of $(k^b_x,k^b_y)$ for the change from reflectance
to transmittance corresponds to a hyperbolic fixed point and its
location can be found conveniently by considering the sizes of the
dispersion curves as discussed in \cite{SCT2015}. The upper fixed
point shown in both parts of Fig. \ref{fig:PhaseSpace4} gives a
threshold value of
$(k^b_x,k^b_y)=(143\mathrm{m}^{-1},80.5\mathrm{m}^{-1})$. We will
also investigate the existence and location of such a threshold
incoming wave direction in the finite difference solution of the
full wave problem described in the next section.

Depending on the angle of incidence, Fig. \ref{fig:PhaseSpace4}
shows that the point of reflection varies from the centre of the
ridge at $x=0$ to slightly towards the flat region to the left. The
smooth curvature model here therefore differs from the discontinuous
curvature models considered in Refs. \cite{L94,N98}, for which the
reflection takes place off-centre at a fixed location only. That is,
for the models presented in Refs. \cite{L94,N98} the reflection
would take place at the location of the jump in the curvature
between the flat region and the cylindrical region. This would be
off-centre in the example here, since the centre point $x=0$
corresponds to the centre of the cylindrical region.

In the next section we consider an equivalent scattering problem to
the one above, but instead using the finite difference model
described in the previous section to numerically solve the full wave
problem. The full wave model will include the phase information
omitted in the pure ray approximation applied in this section, but
at the cost of a computational expense which scales with frequency.
Such a study is feasible up to reasonably high frequencies due to
the one-dimensional setup of the problem derived in \refsec{sec2}.
However, the high frequency purely transmissive behaviour for the
problem here is relatively straightforward to predict, and we find
that both methods may be used for the more interesting mid-frequency
case studied here.

\subsection{Wave scattering finite difference solution}

We restrict the study to bending excitations as in the previous
section. For smooth joints this means that only the bending mode is
active, with negligible conversion to in-plane modes. That is, the
scattering probabilities become
\begin{equation}\label{eq:prob} P\text{(Transmit)}
= |A_{\tau}^b|^2 \quad \text{and} \quad P\text{(Reflect)} =
|A_R^b|^2
\end{equation}
using the scattering amplitudes for bending only. However, for
rapidly changing curvature functions (\ref{eq:curv1}) with very
small $\delta x$ in the interpolation function (\ref{eq:interp}),
mode conversions appeared in the numerics at almost normal
incidence. We defer the study of this case to future work and note
the possibility of using discontinuous joints instead for this case.
Note that we use the term probabilities to describe the coefficients
defined in (\ref{eq:prob}) since for an incoming bending wave of
unit amplitude (\ref{eq:winc}), conservation of energy gives that
$P\text{(Transmit)}+P\text{(Reflect)}=1$. In the full wave picture
considered here these coefficients (\ref{eq:prob}) actually give the
proportions of reflected and transmitted wave energy. However, for
comparison with the ray tracing results in the last section, they
give the probabilities of reflection and transmission for a
particular incoming trajectory.

\begin{figure}[!ht]
\centering
\includegraphics[width=0.9\textwidth]{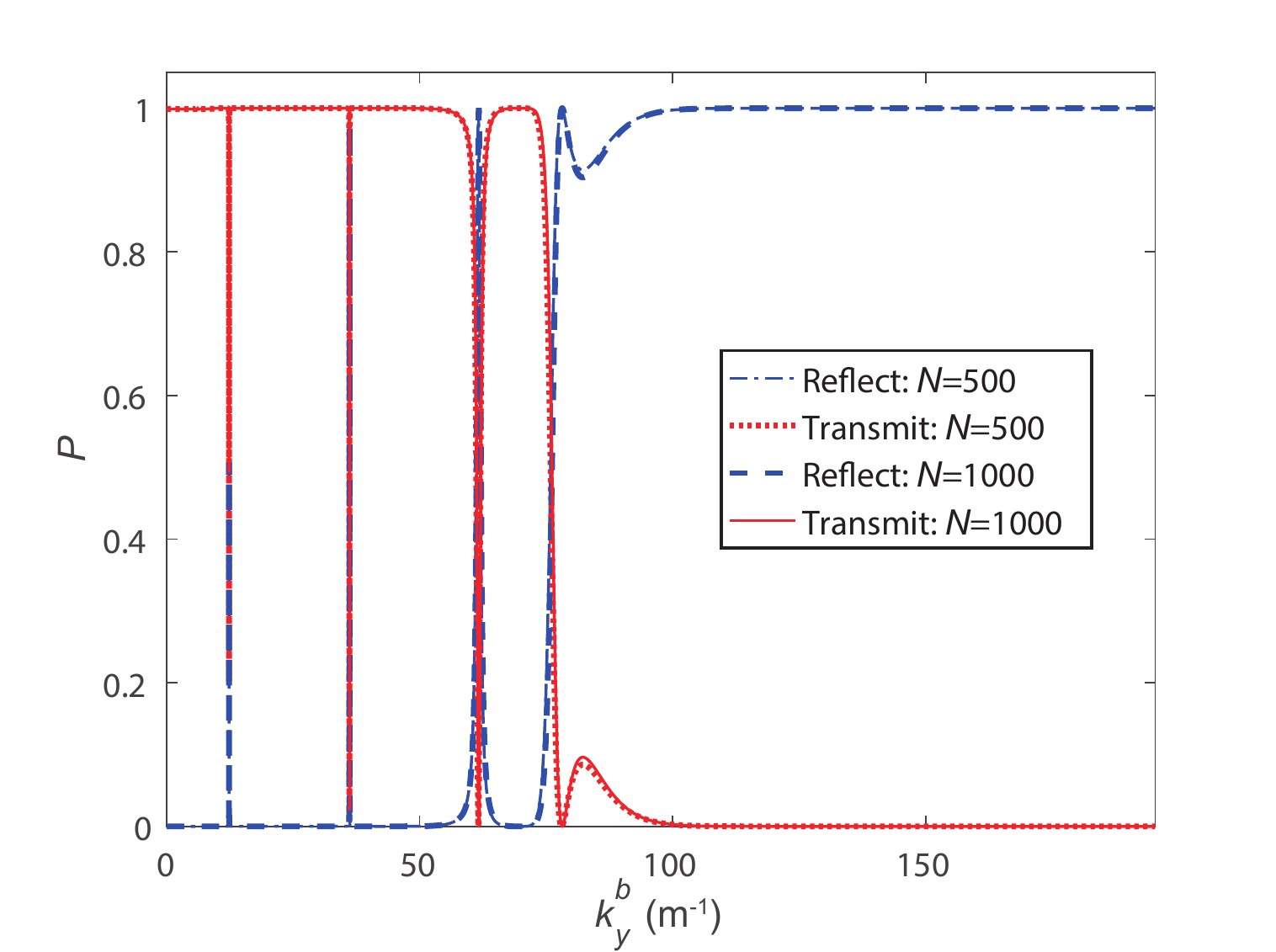} \caption{Reflection and transmission probabilities for the
bending mode as a function of $k_y^b$ plotted on a linear scale. The
plot shows anti-resonances close to $k_y^b=12.3\mathrm{m}^{-1}$,
$k_y^b=36.2\mathrm{m}^{-1}$ and $k_y^b=61.7\mathrm{m}^{-1}$. There
is a switch in the dominant behaviour from transmission to
reflection close to $k_y^b = 78 \mathrm{m}^{-1}$.} \label{fig:Coeff}
\end{figure}
\begin{figure}[!h]
\centering
\includegraphics[width=0.9\textwidth]{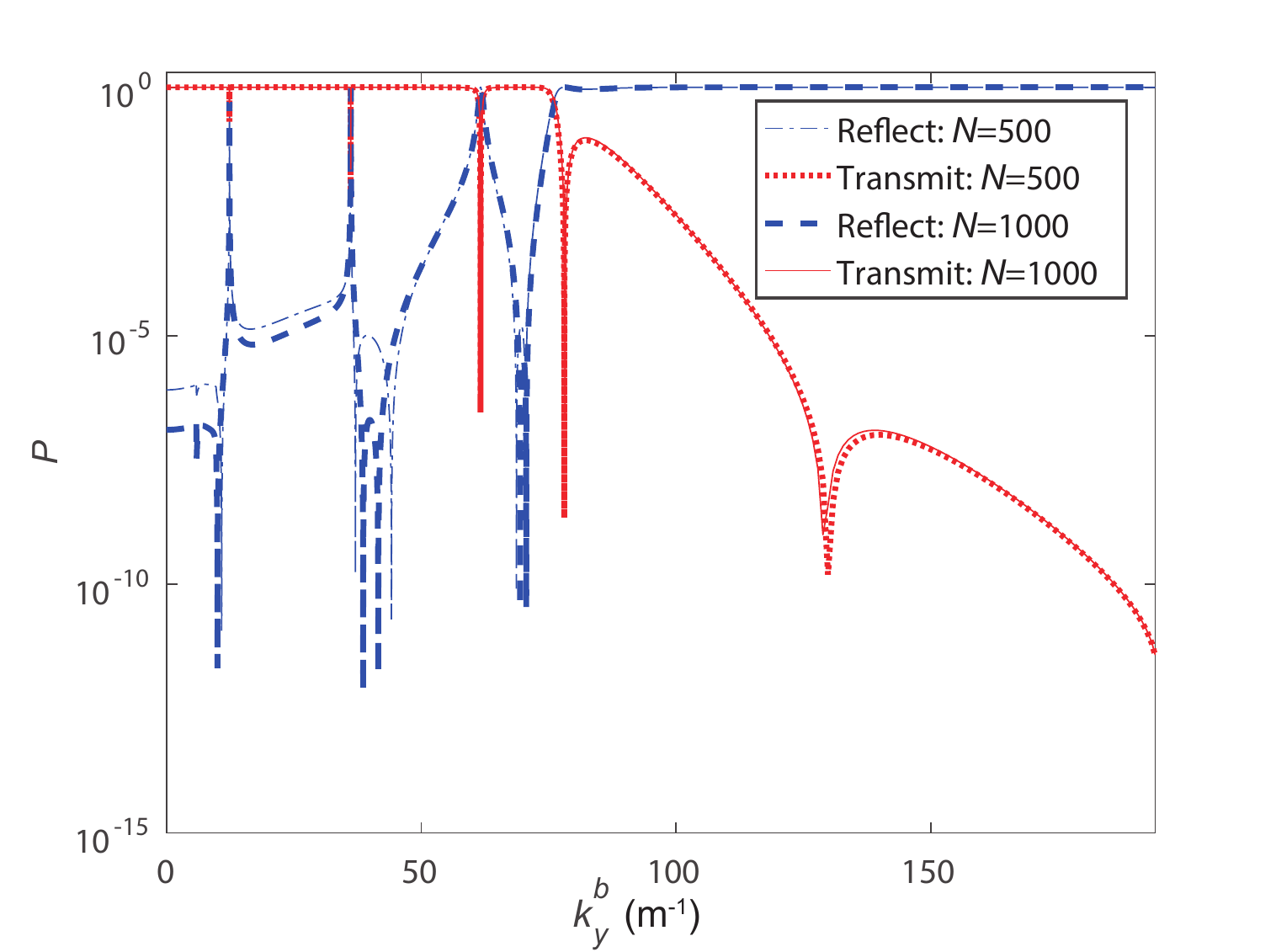}
\caption{Reflection and transmission probabilities for the bending
mode as a function of $k_y^b$ plotted on a logarithmic scale. The
logarithmic scale emphasises the anti-resonance reflection peaks and
the switch from transmission to reflection that were evident from
\reffig{fig:Coeff}.} \label{fig:LogCoeff}
\end{figure}

The dependence of the reflection and transmission probabilities on
the value of the trace wave number $k_y^b$ is shown in
\reffig{fig:Coeff}. The transmission and reflection probabilities
have been computed using the finite difference discretisation
described in \refsec{sec:ScatScheme} for both $N=500$ and $N=1000$
grid points in $\Omega_-$, corresponding to $\Delta x=0.0048384$m
and $\Delta x=0.0024192$m, respectively. The ratio of wavelength to
$\Delta x$ is $6.65$ for the case $N=500$ and hence the rule of
thumb requiring 6 points per wavelength for reliable results
suggests that in both cases our results should be reasonably well
converged. This convergence is also evident from \reffig{fig:Coeff},
since it is difficult to detect differences between the plots for
$N=500$ and $N=1000$.

\reffig{fig:Coeff} shows that the scattering coefficients of the
full wave problem include discrete anti-resonance points (points of
perfect reflection) that were not present in the ray tracing
calculation. In addition, the threshold behaviour from transmission
to reflection has been smoothed in the full wave calculations,
whereas for the ray tracing solution there is a sudden jump from
transmission to reflection corresponding to the location of the
hyperbolic fixed point. Hence, for the calculations in this section
there is a region of $k_y^b$ values where both reflection and
transmission take place at the same time. This region lies roughly
between $k^b_y=73\mathrm{m}^{-1}$ and $k^b_y=95\mathrm{m}^{-1}$, but
one observes an obvious switch from dominant transmission behaviour
to dominant reflection behaviour at a threshold value around $k_y^b
= 78 \mathrm{m}^{-1}$, which is close to the threshold estimate
($k_y^b=80.5\mathrm{m}^{-1}$) from the corresponding ray tracing
calculation. It appears therefore that the asymptotic ray model
slightly over predicts the transmission/reflection threshold
compared with the full wave solution, but still produces a reliable
estimate within around $1\%$ of the total $k_y^b$ range.

A more complete picture appears when the scattering coefficients are
plotted on a logarithmic scale, see \reffig{fig:LogCoeff}. We can
then clearly identify the discrete reflection points in the region
of perfect transmission. The dips or peaks can be arbitrarily close
to zero, or unity, depending on the resolution of the plot. The
discrepancy between the finite difference solution for the
reflection coefficient with $N=500$ and $N=1000$ is also more
evident in this figure since the logarithmic scaling amplifies the
differences at very small values of $P(\text{Reflect})$ below
$10^{-4}$. However, the positions of the anti-resonances and the
switch from transmission to reflection remain in excellent agreement
for both $N=500$ and $N=1000$. Note also that the values of
$P(\text{transmit})$ computed with $N=500$ and $N=1000$ are in good
agreement with one another, even for very small values.

\begin{figure}[!h]
\centering
\includegraphics[width=\textwidth]{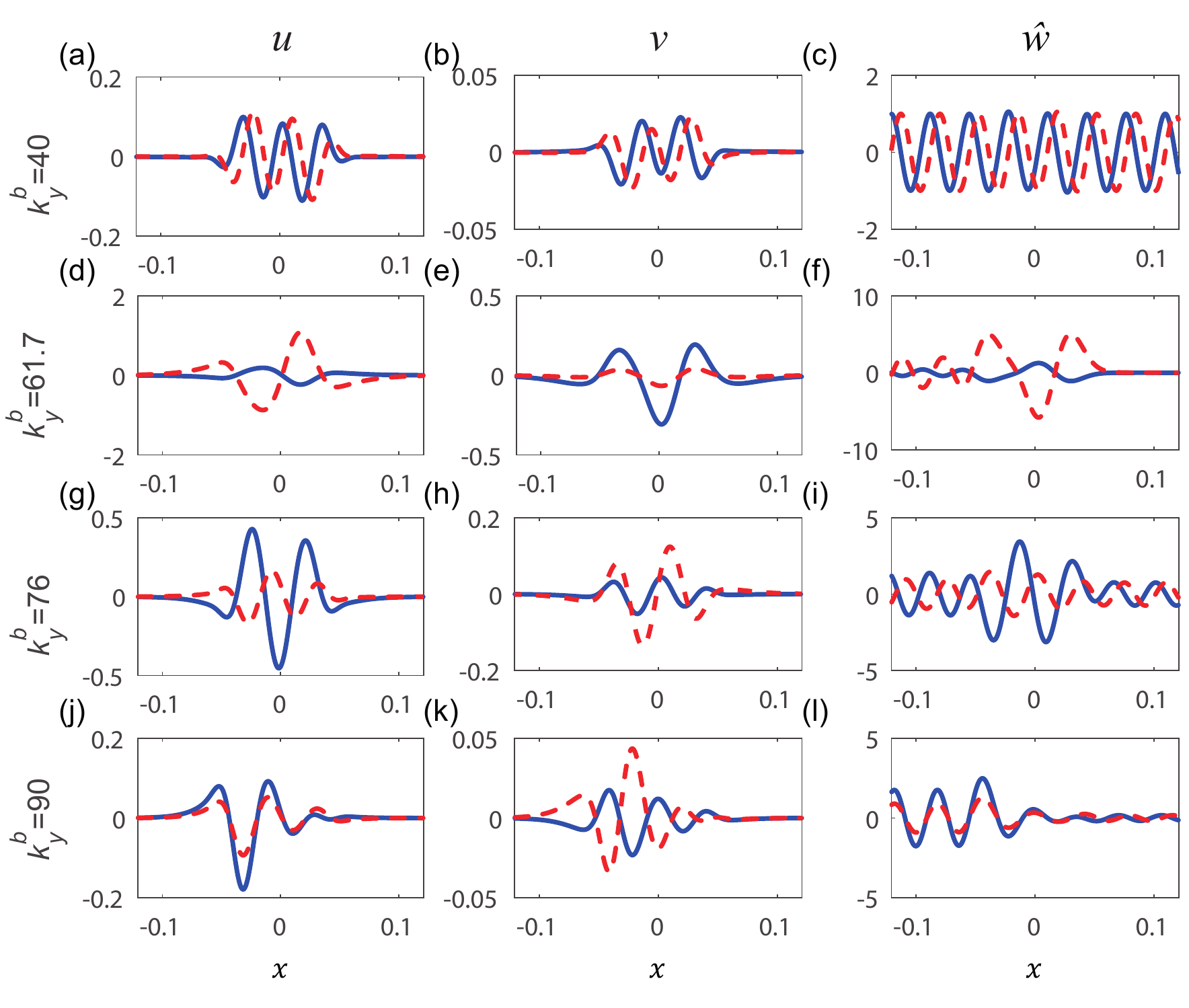}
\caption{Wave function solutions
$\mathbf{\hat{u}}=[u\hspace{3pt}v\hspace{3pt}\hat{w}]$ for
$k_y^b=40\mathrm{m}^{-1}$ (plots (a) to (c)),
$k_y^b=61.7\mathrm{m}^{-1}$ (plots (d) to (f)),
$k_y^b=76\mathrm{m}^{-1}$ (plots (g) to (i)) and
$k_y^b=90\mathrm{m}^{-1}$ (plots (j) to (l)). Each plot shows a
displacement in metres against its corresponding $x$-coordinate,
also in metres. Plots (c), (f), (i) and (l) show the bending mode
$\hat{w}$ which exhibits transmission in plot (c), anti-resonance in
plot (f), mixed reflection and transmission for the threshold region
in plot (i) and reflection in plot (l). The remaining plots show the
in-plane wave solution, which remains confined inside the
cylindrical region in all cases. Solid and dashed lines show the
real and imaginary parts, respectively.} \label{fig:wfMiddle}
\end{figure}

Figure \ref{fig:wfMiddle} shows the behaviour of the wave function
solutions $\mathbf{\hat{u}}=[u\hspace{3pt}v\hspace{3pt}\hat{w}]^\text{T}$
for various values of the trace wavenumber $k_y^b$. The finite
difference computations for these plots were all carried out using
$N=1000$ grid points. Note that there is no mode conversion and so
the in-plane contributions $u$ and $v$ only have support on the
curved region and vanish in $\Omega_+$. In the case of reflection
($k_y^b=90\mathrm{m}^{-1}$) and transmission
($k_y^b=40\mathrm{m}^{-1}$) the expected behaviour is observed. That
is, for the reflective case the bending wave function $\hat{w}(x)$
localises to the left as shown in plot (l) of \reffig{fig:wfMiddle}.
In the ray tracing model one can pinpoint an exact reflection point,
but for the wave problem one observes a decay in amplitude as the
wave enters the cylindrical region rather than a hard wall
reflection.

In the case of transmission shown in plot (c) of
\reffig{fig:wfMiddle}, the bending wave continues through the ridge
without significant deformation from the shape of the initial plane
wave excitation. This is equivalent to the corresponding ray tracing
result. However, if $k_y^b$ is chosen coincide with one of the
anti-resonance peaks in the transmission region that were not
present in the ray tracing result, then the bending wave field is
localised in the central cylindrical region; for example, at
$k_y^b=61.7\mathrm{m}^{-1}$ as shown in plot (f) of
\reffig{fig:wfMiddle}. The bending wave function $\hat{w}(x)$ also
localises in the cylindrical region when $k_y^b$ is chosen in the
threshold region as shown in plot (i) of \reffig{fig:wfMiddle},
which shows a plot of $\hat{w}(x)$ at $k_y^b=76\mathrm{m}^{-1}$.

\section{Conclusions}\label{sec:conclu}

We have investigated bending wave scattering across a smoothed plate
and quarter cylinder configuration in the interesting mid-frequency
case, close to the ring frequency for the cylinder. Results have
been obtained in a simplified and effectively one-dimensional
setting using both a high-frequency ray-tracing approximation and a
finite difference discretisation of the equivalent full-wave
problem. Previous studies on plate and cylinder connections have
concentrated on joints having discontinuous curvature where
reflections can occur, however the smoothed joints considered here
also give rise to reflections within the cylindrical part of the
structure. Furthermore, the ray tracing calculations suggest the
existence of a threshold incident wave direction which separates
waves or rays that exhibit reflective or transmissive behaviour.
This threshold direction is also evident from the smoothed
transition between reflection and transmission observed in the full
wave calculations. Hence, relatively simple scattering laws can be
employed to model the propagation of structure-borne noise in
shells, and ultimately in built-up structures containing thin shell
components. The full wave solution shows that in addition to the
switch from transmission to reflection as the incident wave
direction becomes increasingly oblique, there are also
anti-resonances giving rise to perfect reflection at a discrete set
of directions where transmission would typically be expected. In
these cases the wave functions appear as trapped modes that localise
in the cylindrical part of the configuration.

There are several avenues for further research, including extensions
to multiply-curved and fluid-loaded shells. In addition, the
scattering laws could be incorporated within computed aided
engineering simulations of built-up structures via wave methods,
such as the wave and finite element method or dynamical energy
analysis.

\section*{Acknowledgments}
Support from the EU (FP7-PEOPLE-2013-IAPP grant no.~612237 (MHiVec))
is gratefully acknowledged. We also wish to thank Dr Gregor Tanner
for stimulating discussions and Dr Jonathan Crofts for carefully
reading the manuscript.

\bibliographystyle{plain}

\appendix

\section{Tensors and differential geometry}\label{sec:Tensors}

Several physical theories are concisely written in tensor form
\cite{F72}. The tensor formalism has also proved to be useful in
continuum mechanics. The components of tensors are with respect to a
given choice of coordinates $(x^i)$. A change of coordinates from
$(x^i)$ to $(\tilde{x}^{i})$ leads to expressions for components in
the new coordinates in relation to those of the old. For example,
given a tensor of type $(1,2)$ (that is, with one superscript and
two subscript indices) the transformation takes the form
\begin{equation}
\tilde{T}_{a b}^{c} = \frac{\partial \tilde{x}^c}{\partial x^{c'}}
\frac{\partial x^{a'}}{\partial \tilde{x}^{a}} \frac{\partial
x^{b'}}{\partial \tilde{x}^b} T^{c'}_{a' b'}
\end{equation}
where Jacobians and inverse Jacobians have been used for the change
of coordinates. We have adopted the important summation convention
for repeated indices; this indicates that a summation is to take
place over the repeated index (unless otherwise stated).

The metric tensor $g_{\alpha \beta}$, known as the first fundamental
form, measures distances via
\begin{equation} \mathrm{d}s^2 = g_{\alpha
\beta} \mathrm{d}x^\alpha \mathrm{d}x^\beta.
\end{equation} The inverse of the metric
tensor $g^{\alpha \beta}$ obeys
\begin{equation}
\delta_{\alpha}^{\beta} = g_{\alpha \gamma} g^{\gamma \beta}
\end{equation}
and is used to alter the type of a tensor. For example, we can
introduce a new tensor of type (1,1)
\begin{equation} T^\gamma_\alpha=T_{\alpha \beta} g^{\beta \gamma}
\end{equation}
from a tensor of type (0,2) by raising the indices of the latter
tensor. Likewise $g_{\alpha \beta}$ lowers indices.

The  directional derivative  in arbitrary coordinates  is
generalized to the covariant derivative $D_\alpha$. This derivative
may be introduced from an embedding using projection of gradients
\cite{N98, C76} or intrinsically with quantities only related to the
curved space  itself \cite{F04}. For example, in a coordinate basis,
the covariant derivative of a tensor of type (1,1) becomes
\begin{equation} D_\alpha u_\beta^\epsilon= \partial_\alpha
u_\beta^\epsilon - \Gamma_{\alpha \beta}^\gamma u_\gamma^\epsilon  +
\Gamma_{\alpha \gamma}^\epsilon u^\gamma_\beta \end{equation} with
\begin{equation} \Gamma_{k l}^\mathrm{i}= \frac{1}{2} g^{m} (\partial_l
g_{mk} +\partial_k g_{m l} -
\partial_m g_{k l}) \end{equation} the Christoffel symbols.

\section{Parameter values for the numerical studies}
\label{sec:Parameters} The computations throughout this work were
done using the following parameter choices \cite{WHB90,RN96}:
\begin{itemize}
\item $R=0.055$ m\vspace{-1mm}
\item $h=5.3\times10^{-4}$ m\vspace{-1mm}
\item $x^*=0.0432$ m\vspace{-1mm}
\item $\delta x =0.0144$ m\vspace{-1mm}
\item $E=1.95\times10^{11}$ Pa\vspace{-1mm}
\item $\rho=7700$ $\mathrm{\mathrm{kg}/\mathrm{m}^3}$\vspace{-1mm}
\item $\nu=0.28$\vspace{-1mm}
\item $\omega=9742\pi$ Rad/s
\end{itemize}
\end{document}